\documentclass[aps,pre,twocolumn,groupedaddress,showpacs,amsmath,amssymb,merge]{revtex4}
\usepackage{graphicx}

\begin{document}

%Title of paper
\title{Modulation effects within the mean-field theory of electrolyte solutions}

\author{Jonathan Landy}
\email[]{landy@physics.ucla.edu}

\affiliation{Department of Physics and Astronomy, University of
California Los Angeles, Los Angeles, CA 90095-1547, USA}

\date{\today}

\begin{abstract}
The consequences of source charge and surface modulation are studied within the framework of the Poisson-Boltzmann theory of electrolyte solutions.  Through a consideration of various examples, it is found that inherent modulation can lead to both like-charge attraction and overcharging effects.         
\end{abstract}


\maketitle

\section{Introduction}

As in any conductor, the potential due to a source distribution is screened within an electrolyte solution.   The screening charge carriers, or counterions, in these classical systems are ionized atoms or molecules. The qualitative result of the screening is typically only to reduce the distance out to which the potential from a source distribution is significant.  However, experiments have shown that multivalent counterions can allow for more exotic behavior \cite{Gel:00}.  For example, over-efficient screening due to multivalent counterions can sometimes result in the charge reversal of a molecule or substrate.  In a related effect, multivalent counterions have been observed to mediate attractions between similarly charged macromolecules.  Study of these counterintuitive effects is pertinent to the understanding of many biological and colloid systems, in which molecular conformations and interactions are often sensitively controlled by counterion concentration.

The Poisson-Boltzmann (P-B) mean-field theory provides one of the simplest descriptions of electrolyte response and often allows for accurate modelling of these systems.  However, it has been proven for mirror-symmetric geometries that like-charge attraction is impossible for a pair of identical molecules within this theory \cite{Sad:00}.  This has bolstered the general opinion that the physical mechanisms behind like-charge attraction and overcharging cannot be understood within a mean-field description of electrolytes.  Most recent works have thus attempted to understand these observations through a consideration of fluctuation corrections to the mean-field response \cite{Oos:68,Man:69,Ray:94,Bar:96,Rou:96,Gro:97,Ha:97,Kor:97:1,Pod:98,Lev:992,Shk:99,Ngu:00}.  Certainly, for homogeneous systems, it seems clear that fluctuations provide the only possible mechanism for bringing about the charge correlations necessary for these effects.  This conclusion does not immediately follow for heterogeneous systems, however, since static modulation allows for the introduction of some inherent charge correlation.  Indeed, studies have shown that heterogeneity can play an important role in these systems.
For example, numerical simulations have shown that modulation of molecular geometry can significantly affect interactions \cite{All:04}.   In addition, it has been shown perturbatively that within the P-B theory, charge condensation onto a surface in an electrolyte solution may be increased when the surface charge contains certain limiting forms of modulation  \cite{Luk:02}.  This extra condensation can in turn lead to a decrease in the osmotic pressure between two surfaces, demonstrating that mean-field effects can at least help to facilitate both overcharging and like-charge attraction \cite{Luk:022}.

It is the purpose of the present paper to further explore the effects of inherent, quenched modulations within electrolyte solutions.  Fluctuation effects are neglected and the electrolyte response is considered within the P-B formalism.  Significantly, it is shown that inherent modulations alone, can, in fact, allow for both like-charge attraction and overcharging mechanisms.  The key to like-charge attraction within the P-B theory is the breaking of mirror symmetry, which was assumed to hold in Ref.~\cite{Sad:00}.  A simple example is given which demonstrates this point.  Heterogeneity-induced condensation is studied using a perturbation approach similar in spirit to that considered in Ref.~\cite{Luk:02}.  The present approach consists of expanding about solutions to the linearized P-B equation and allows for the consideration of additional charge modulation limits, boundary surface modulations, and the effects of having more than one species of screening ion.  Various examples are considered in order to provide a brief survey of these different set-ups.  Of particular interest is the counterion concentration dependence exhibited by systems containing multiple species of ion.  In principle, this dependence should often allow one to determine whether modulation is a significant contributor to overcharging in any particular experimental observation.

An outline of this paper is as follows.  In section II the P-B theory is reviewed and solutions to the linearized P-B, or Debye-H\"{u}ckel \cite{Deb:23}, equation are presented.  Charge and surface modulations are considered in sections III and IV, respectively, while section V contains concluding remarks.

\section{The Poisson-Boltzmann equation}
If one assumes that the mobile charge distribution in an electrolyte solution is given by a sum over ion species of Boltzmann factors multiplying bulk charge densities, then the Poisson equation reads

\begin{eqnarray}\nonumber \label{PBE}
\nabla^2 \phi &=& - 4 \pi \sum_i e^{-q_i \phi/k T} q_i n_i - 4 \pi \rho_s \\
&\approx & \gamma^2 \phi - 4 \pi \rho_s.
\end{eqnarray}
Here, $\phi$ is the electrostatic potential, the $q_i$ and $n_i$ are the charge and bulk densities of screening species $i$, $k T$ is the thermal energy, $\rho_s$ is the source distribution, and $\gamma \equiv l_D^{-1}$ is the inverse Debye length  \cite{Han:00}.  The first equation above is the full P-B equation.  The linearized approximation is written out in the second line.  Solutions to these equations provide mean-field approximations to the potentials of physical systems.  

It is possible to derive the P-B equation in a rigorous manner which clarifies the physical approximations and assumptions required to arrive at this expression.  In the weak potential limit, when the linearized form applies, it has been shown that the equation is exact, while errors result at each non-linear order \cite{Ons:33,Kir:54}.  However, recent simulations have shown that a partial cancellation of two neglected effects, counterion correlations and finite ion sizes, can result in approximate agreement between the P-B solutions and experiments \cite{Des:00}.  Scattering experiments have also shown that the full non-linear solutions are in reasonable agreement with observed charge distributions when only a single screening ion species is present \cite{Das:03}, though some extra divalent condensation was observed when both monovalent and divalent ion species were present \cite{And:04}.  These results justify use of the non-linear equation, but the solutions should be considered qualitatively accurate only outside the innermost, Stern layer surrounding highly charged distributions.

Solutions to the inhomogeneous, linearized P-B equation will be needed throughout this paper.  The Poisson sum rule provides a convenient method to determine the potential of a lattice of point charges, from which one can obtain the potentials of more general distributions through superposition.  For an arbitrary individual particle potential function $f$, the rule states
\begin{eqnarray}
\label{Poisson}
\sum_{\textbf{R}} f(\textbf{r}-\textbf{R}) = \frac{1}{V} \int \sum_{\textbf{G}}f(\textbf{r}-\textbf{R}) \exp(i \textbf{G}\cdot \textbf{R}) d^n R.
\end{eqnarray}
Here, $n$ is the dimension of the lattice, $V$ is the $n$-dimensional volume per unit cell of the direct lattice, the vectors $\textbf{R}$ are the direct lattice vectors of the distribution, and the vectors $\textbf{G}$ are the reciprocal lattice vectors of the distribution \cite{Zim:72}.  Shifting the origin of integration in Eq.~(\ref{Poisson}) to the projected position of the observation point $\textbf{r}$ immediately gives formal integral representations for the Fourier coefficients of the potential.

To obtain the potential from a lattice of Coulomb point charges which are linearly screened, one may plug in the Yukawa potential function $f = q \exp[-\gamma r]/r$, which is the solution to the linearized P-B equation with point charge source $q$.  The resulting expressions for the potentials of one, two, and three-dimensional Yukawa lattices are \cite{Cra:87}
\begin{eqnarray}
\label{oneds}
\Phi_{1,Y} &=& 2 \lambda \sum_{\textbf{G}}  K_0((|\textbf{G}|^2 + \gamma^2)^{1/2}d)\exp[i \textbf{G}\cdot \textbf{r}], \\ 
\label{twods}
\Phi_{2,Y} &=& 2 \pi \sigma \sum_{\textbf{G}} \frac{ \exp[i \textbf{G}\cdot \textbf{r}-(|\textbf{G}|^2 + \gamma^2)^{1/2}d]}{(|\textbf{G}|^2 + \gamma^2)^{1/2}}, \\
\label{threeds}
\Phi_{3,Y} &=&  4 \pi \rho \sum_{\textbf{G}} \frac{1}{|\textbf{G}|^2+\gamma^2} \exp[i \textbf{G}\cdot \textbf{r}],
\end{eqnarray}
where $\lambda$, $\sigma$, and $\rho$ are the average charge densities of the lattices.  For large arguments, the modified Bessel function $K_0$ above may be expanded as \cite{Abr-72}
\begin{eqnarray}
K_0(z) \sim \sqrt{\frac{\pi}{2 z}}e^{-z} \sum_{j=0}^{\infty}\frac{(-1)^j \prod_{k=0}^{j}(2k+1)^2}{j! (8z)^j}.
\end{eqnarray}
It follows that each component will be exponentially damped with both the frequency and the distance from the distribution for one and two-dimensional systems.  To relate these expressions to their unscreened analogs, one need only replace $(|\textbf{G}|^2 + \gamma^2)^{1/2}$ by $|\textbf{G}|$ in the non-zero frequency components and then also replace the zero frequency component by the appropriate continuous charge distribution expression.  This is given by $-2 \lambda \log r$ for a linear distribution, for example.

\section{Source charge modulation}
\begin{figure}\scalebox{0.68}
{\includegraphics{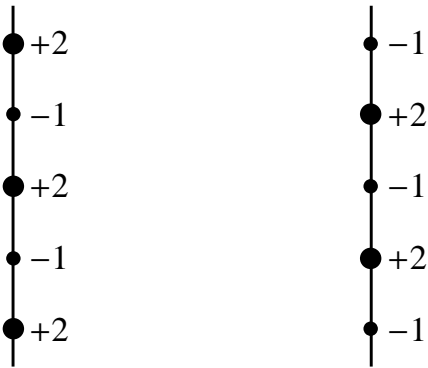}}
\caption{\label{fig:altlines}  Portions of two long, identical polyelectrolytes are shown.    At large distances they repel but at short distances they attract.}
\end{figure}

Consider now the charge modulated system depicted in Fig.~\ref{fig:altlines}.  Two model polyelectrolytes are shown, each of which consists of alternating point particles of charge $+2$ and $-1$ fixed on a rigid backbone.  The period of the modulation is taken to be $1$.  Note that each polyelectrolyte has a net charge of $+1$ per period.  If the two polyelectrolytes are shifted with respect to one another by half of a period, as shown in the figure, then they will attract one another when their separation is small, demonstrating that mean-field like-charge attraction is possible.  

\begin{figure}\scalebox{0.9}
{\includegraphics{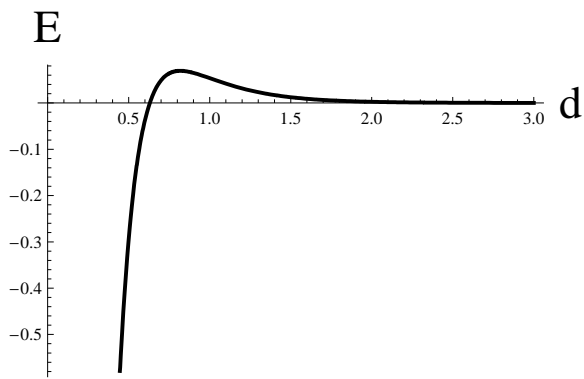}}
\caption{\label{fig:linesenergy} A plot of the approximate interaction energy versus separation for the model polyelectrolytes.}
\end{figure}

One can approximate the interaction energy of this system in the Debye-H\"{u}ckel limit by superposing the potential for an infinite line of $-1$ charges with that for a line of $+2$ charges, each obtained by plugging into Eq.~(\ref{oneds}), and truncating the series after two terms.  
The resulting approximate interaction energy per period of the second polyelectrolyte is given by

\begin{eqnarray} \label{energy:approx}
E(d) \approx 2 \mbox{K}_0[\gamma d] - 36 \mbox{K}_0[\sqrt{(2 \pi)^2 + \gamma^2}d],
\end{eqnarray}
where $d$ is the separation distance between the two polyelectrolytes.  A plot of this function is shown in Fig.~\ref{fig:linesenergy} for $\gamma =3.0$.  Attraction is indeed observed at small separations while repulsion is observed at large separations.  This form of attraction is not salt mediated, but as $\gamma$ is increased, the repulsive effect at large distances is reduced.  This trivial example demonstrates that like-charged molecules can be attracted to one another even when the electrolyte is considered at the mean-field level.  Note, however, that if the model polyelectrolytes are arranged in a mirror-symmetric orientation, as was explicitly assumed in Ref.~\cite{Sad:00}, then the coefficient of the second term in Eq.~(\ref{energy:approx}) would also be positive.  This is consistent with the conclusion that identical molecules cannot be attracted to one another, within the P-B formalism, when they are arranged in a mirror-symmetric orientation.

We now turn to a consideration of charge modulation-induced condensation.  As mentioned above, this was previously studied in Refs.~\cite{Luk:02} and \cite{Luk:022}.  These previous works considered solutions to the full non-linear P-B equation for systems containing a planar source distribution and only a single species of screening ion.  Solutions were obtained up to second order in the parameters $\epsilon(\vec{q})/\bar{\sigma}$, for all $\vec{q} \not = 0$.  Here, $\bar{\sigma}$ was the average charge density of the planar source distribution and the $\epsilon(\vec{q})$ were the amplitudes of the oscillatory components of the source distribution's Fourier decomposition.  Thus, these solutions were valid in the limit of relatively weak modulation relative to net surface charge.  Interestingly, these solutions indicated that each non-zero Fourier component independently led to an increase of screening charge condensation at second order in perturbation theory.  This universal result was also shown to hold for long wavelength modulations.  Here it is shown that a similar conclusion also holds true in the general weak potential limit.  

\begin{figure}\scalebox{0.9}
{\includegraphics{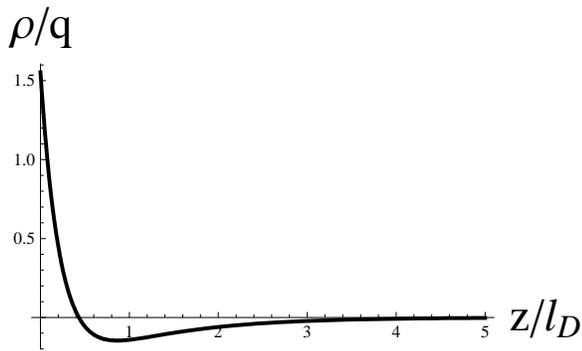}}
\caption{\label{fig:condensation} A plot of the one component fluid's second order charge condensation versus $z$, the distance from the plane.  Here $G = 2 \gamma = 2/l_D$.  The condensation is particularly localized and then screened at larger distances.  The localization increases exponentially with the magnitude of $G$.  The total integrated charge from $z = 0$ to $\infty$ is zero for each component, as expected. }
\end{figure}

For simplicity, we consider first a system which contains only one species of counterion.  The counterions are taken to have charge $q$ and bulk density $n_0$, so that the P-B equation reads

\begin{eqnarray}
\nabla^2 \phi &=& -4 \pi q n_0 e^{-q \phi/ kT} - 4 \pi \rho_s \\
&\approx & -4 \pi q n_0 (1 - \frac{q}{kT}\phi + \frac{1}{2}(\frac{q}{kT})^2\phi^2) - 4 \pi \rho_s. \nonumber
\end{eqnarray}
Here, $\rho_s$ is a perturbing planar charge distribution.  The first term in the above Boltzmann factor expansion corresponds to the uniform background charge density.  This does not effect an electric field and so may be dropped for now.  We assume that the potential can be expanded in a series as $\phi =  \phi_1  + \phi_2 + ...$, where the term $\phi_k$ is of order $\rho_s^k$.  Collecting like-powers of $\rho_s$ then gives the system of equations
\begin{eqnarray}\label{pert:series} \nonumber
\nabla^2 \phi_1 &=& \frac{4 \pi q^2 n_0}{kT} \phi_1 - 4 \pi \rho_s \\ \nonumber
\nabla^2 \phi_2 &=& \frac{4 \pi q^2 n_0}{kT} \phi_2 - \frac{2 \pi q^3 n_0}{(k T)^2}\phi_1^2 \\
& ...& 
\end{eqnarray}
The solution to the first equation above may be obtained from our earlier expressions if the distribution $\rho_s$ is expanded in a Fourier series as
\begin{eqnarray}
\rho_s = \frac{1}{2 \pi} \sum_{\textbf{G}}A_{\textbf{G}}\exp[i \textbf{G} \cdot \textbf{r}].
\end{eqnarray}
Note that while extra, oscillatory condensation results at this first order, it averages to zero when integrated over the modulation directions.  The source term for $\phi_2$ is proportional to the square of the first order potential, however, and a non-zero condensation will result at this order.   To obtain the net condensation, one need only consider the average charge density above the surface.  The equation for $\phi_2$ is therefore averaged over $x$ and $y$ to obtain
\begin{eqnarray}\nonumber
( \frac{\partial^2}{\partial z^2} - \gamma^2)\overline{\phi_2 } 
&=& - \frac{\gamma^2}{2}(\frac{q }{ k T})\sum_{\textbf{G}} |A_{\textbf{G}}|^2 \frac{e^{-2 (G^2 + \gamma^2)^{1/2}z}}{G^2 + \gamma^2},\\
\end{eqnarray}
an ordinary differential equation which may be solved trivially.  The solution is
\begin{eqnarray}
\label{planesol1}
\lefteqn{\overline{\phi_2} =\gamma( \frac{ q}{k T})\sum_{\textbf{G}}\frac{|A_{\textbf{G}}|^2}{\sqrt{G^2 + \gamma^2}(4 G^2 + 3 \gamma^2)}} \\ & \times & \{\exp[-\gamma z] - \frac{\gamma}{2 \sqrt{G^2 + \gamma^2}}\exp[-2 \sqrt{G^2 + \gamma^2}z]\}. \nonumber
\end{eqnarray}
A plot demonstrating the resulting $z$ dependence of a typical term of the second order average charge density is shown in Fig.~\ref{fig:condensation}.  The total averaged second order counterion charge density at the surface is given by the following sum over modulation components:
\begin{eqnarray}
\label{planesol2}
\overline{\rho_2}(z=0) &=&-\frac{1}{4 \pi} \frac{\partial^2}{\partial z^2} \overline{\phi_2} \\ \nonumber
&=& q n_0(\frac{q}{kT})^2 \sum_{\textbf{G}}\frac{|A_{\textbf{G}}|^2}{4 G^2 + 3\gamma^2}(2  - \frac{\gamma}{\sqrt{G^2 + \gamma^2}}).
\end{eqnarray}
This demonstrates that indeed there is extra condensation proportional to the square of the amplitude for each mode of the modulation, consistent with the long wavelength and small modulation results of Refs.~\cite{Luk:02} and \cite{Luk:022}.  The present analysis extends the conclusion that universal modulation-induced condensation is expected at second order to all relative values and wavelengths of modulation.  In particular, the result applies to systems which have little or no net charge but large modulations.  In systems such as these, the modulation-induced condensation could easily result in overcharging.  This demonstrates that mean-field overcharging is possible in certain limits.

A single component electrolyte is somewhat non-physical in that the solution contains charges of only one sign and thus cannot be net neutral.  Accordingly, it is of interest to consider systems which contain multiple species of ions.  This would be difficult when considering solutions to the full P-B equation but the present solution technique is readily applied to such systems.  In a general, neutral electrolyte solution, the potential satisfies the approximate equation
\begin{eqnarray}
\nabla^2 \phi \approx \gamma^2 \phi -\zeta \phi^2 - 4 \pi \rho_s.
\end{eqnarray}
Here, the coefficients $\gamma^2$ and $\zeta$ are now given by sums over ions species as
\begin{eqnarray}
\gamma^2 &=& \frac{4 \pi}{k T}\sum_{i} q_i^2 n_i \\ \label{zetaeq}
\zeta &=& \frac{2 \pi}{(k T)^2} \sum_{i} q_i^3 n_i.
\end{eqnarray}
Charge neutrality requires $\sum q_i n_i = 0$, if the charge of the polyelectrolyte itself is neglected, for simplicity.  Solving the resulting system of equations as in the above gives the following for the average electrolyte charge density at the surface: up to second order,
\begin{eqnarray}\nonumber \label{multspecies}
\lefteqn{\overline{\rho}(z=0) = \rho_0 } \\ &&+ \frac{\zeta}{2\pi}\sum_{\textbf{G}\not = 0} \frac{|A_{\textbf{G}}|^2}{4 G^2 + 3 \gamma^2}(2 - \frac{\gamma}{\sqrt{\gamma^2 + G^2}}),
\end{eqnarray}
where $\rho_0$ is the charge density which would appear for a uniform distribution of charge.  

It is interesting to consider some particular examples.  Consider first a net neutral electrolyte system containing screening charges of both signs and equal magnitudes.   For this system, the quadratic terms in the expansion of the Boltzmann factors will cancel and no second order potential or condensation will result.  This implies that any extra condensation would have to occur at higher order for such systems and the resulting effect should be much weaker.  The quadratic term in the expansion is retained, however, for a net neutral system which contains charges of different sign and different valence.  For example, if the positive counterions have charge $q_1$ with bulk density $n_{0}$ and the negative counterions have charge $-q_2$ with bulk density $\frac{q_1}{q_2}n_{0}$, then the resulting second order coefficient is given by 

\begin{eqnarray}
\zeta = \frac{2 \pi q_1 n_0}{(k T)^2}(q_1^2 - q_2^2).
\end{eqnarray}
The result is that extra charge condensation still occurs at the lowest non-linear order for a net neutral system, provided the free charges have different valences.  Note that the net sign of the resulting modulation-induced condensation does not depend on $A_0$.   Extra condensation occurs for both species of ion, and the sign of the net condensation depends only on which species dominates.  This, in turn, is determined by the sign of the quadratic term in the expansion of the P-B equation; if $\zeta$ is large and positive, the resulting condensation will be large and positive, and vice versa.  This obviously holds for any distribution of counterion species.  Further comments on this point appear in the discussion section of the paper.

One and three-dimensional charge modulated distributions may be analyzed in a similar manner.  For a three-dimensional lattice, the concept of condensation is not well-defined.  In the one-dimensional case, however, an analogous condensation to that discussed above for surface distributions also occurs.  Each modulation component leads to additional condensation at second order, the sign of which depends only on the sign of the quadratic term in the expansion of the P-B equation.  A consideration of similar surface modulation effects is contained in the following section.

\section{Surface modulation}
In the above, we saw how charge modulation can allow for oscillatory potential components that lead to a net charge condensation at non-linear orders.  A modulated surface can also allow for the introduction of potential modulation and similar condensation effects must naturally result.  It is worthwhile to consider some examples, and two are presented here which are susceptible to perturbation expansions.  These examples are of interest, not because they accurately model physical systems, but because they demonstrate the mechanisms involved for different boundary conditions and geometries.

A constant potential boundary condition problem will be considered first, where the surface takes the form
\begin{eqnarray}
Z(x,y) = \beta \sum_{\textbf{G} \not =0}A_{\textbf{G}}e^{i \textbf{G}\cdot \textbf{r}}.
\end{eqnarray}
Here $\beta$ is a small constant and the $A_{\textbf{G}}$ are assumed to be $O(\beta^0)$.  At linear order, the system of equations to be solved is
\begin{equation}
\begin{cases} (\nabla^2 - \gamma^2) \phi_1 =0 & \text{if $z \geq Z(x,y)$,}
\\
\phi_1 = 1 &\text{if $z=Z(x,y)$,}
\\
\phi_1 \rightarrow 0 & \text{as $z \rightarrow \infty$,}
\end{cases}
\end{equation}
and the term $\phi_1$ is assumed expandable as
\begin{eqnarray}
\phi_1 = \sum_{k=0}^{\infty}\phi_{1,k}\beta^k.
\end{eqnarray}
Each of the $\phi_{1,k}$ satisfy the Debye-H\"{u}ckel equation individually and may be determined iteratively by expanding the boundary condition as
\begin{eqnarray}\label{2dbcs}
\nonumber
1 &=& \phi_1(z=Z) \\ &=& \sum_{n=0}^{\infty}(\sum_{\textbf{G}}A_{\textbf{G}}e^{i \textbf{G}\cdot \textbf{r}})^n \frac{\beta^n}{n!}\frac{\partial^n \phi_1}{\partial z^n}|_{z=0}.
\end{eqnarray}
Equating coefficients at order $\beta^k$ on both sides of Eq.~(\ref{2dbcs}) generates the boundary conditions for the $\phi_{1,k}$ at the simple surface $z=0$.  Solving up to second order in $\beta$ gives
\begin{eqnarray} \label{sur1}
\phi_{1,0} &=& \exp[-\gamma z] \\
\phi_{1,1} &=& \sum_{\textbf{G}}\gamma A_{\textbf{G}} e^{i \textbf{G}\cdot \textbf{r}}\exp[- \sqrt{G^2 + \gamma^2}z] \\ 
\overline{\phi_{1,2}} &=& \sum_{\textbf{G}} |A_{\textbf{G}}|^2 (\gamma \sqrt{G^2+\gamma^2} - \frac{\gamma^2}{2!})\exp[- \gamma z].
\end{eqnarray}
The last term has once again been averaged over the modulation directions to simplify the calculation.  As in the charge modulation discussion, each modulation component is observed to independently result in extra second order condensation.  Here, however, the surface modulations generate the extra charge condensation at linear order in the P-B theory.  Since the effect takes place at linear order, the sign of the condensation does not depend on the sign and distribution of the counterions, but instead is always opposite to that of the surface potential.  

The first order solution above can easily be plugged back into the non-linear P-B equation to obtain the next order term in the solution.  Although this won't be done here, we note that the non-linear terms in the P-B expansion can result in extra condensation at the same order in the potential and in $\beta$ as that obtained above.  These terms, however, are sensitive to the sign of $\zeta$.  In particular, if $\zeta = 0$, the linearized P-B solution will be correct to lowest order.  Alternatively, the higher order solutions can result in charging of the wrong sign, or negative screening, if $\zeta$ is chosen appropriately.

We now consider a nearly cylindrical geometry with boundary conditions which are more applicable to biological molecules.  The set-up consists of a line of charge of unit linear charge density surrounded by a dielectric material of shape

\begin{eqnarray}\label{Rdef}
R= 1 + \beta \cos (G z).
\end{eqnarray}
Outside the dielectric near-cylinder the potential is assumed to satisfy the linearized P-B equation. The following system of equations specifies the unique solution for the potential:
\begin{equation}
\begin{cases} (\nabla^2 - \gamma^2) \phi =0 & \text{if $r \geq R(z)$,}
\\
\nabla^2 \phi = - \frac{4 \pi}{\epsilon_i}\rho &\text{if $r<R(z)$,}
\\
\phi \rightarrow 0 & \text{as $r \rightarrow \infty$,}
\\
\phi_i = \phi_o & \text{at $r = R(z)$,}
\\
\textbf{D}_i\cdot \hat{\textbf{n}} =\textbf{D}_o\cdot \hat{\textbf{n}} & \text{at $r = R(z)$.}
\end{cases}
\end{equation}
Here, the subscripts $i$ and $o$ stand for inside and outside, respectively, $\epsilon$ is the dielectric coefficient, $\textbf{D}$ is the electric displacement, $\hat{\textbf{n}}$ is the outward normal to the near cylinder's surface, and $\rho = \delta(r)/2 \pi r$.

The boundary conditions may be expressed at the surface $r=1$ in a similar manner to that presented in Eq.~(\ref{2dbcs}).  For example, the boundary condition relating the equality of the normal component of the electric displacement inside and outside the dielectric-solution interface takes the form

\begin{eqnarray}\nonumber 
\lefteqn{\epsilon_i  \sum_{k=0}^{\infty} \frac{( \beta \cos (G z)) ^k}{k!} \frac{\partial^k \nabla \phi_i \cdot \hat{\textbf{n}}}{\partial r^k}\mid_{r=1}}  \\ &=&
\epsilon_o  \sum_{k=0}^{\infty} \frac{( \beta \cos (G z)) ^k}{k!} \frac{\partial^k \nabla \phi_o \cdot \hat{\textbf{n}}}{\partial r^k}\mid_{r=1},
\end{eqnarray}
where the normal is given explicitly by
\begin{eqnarray} 
\label{nhateqn}
\hat{\textbf{n}} = \frac{\textbf{r}}{\sqrt{1 + \beta^2 G^2 \sin^2(G z)}} + \frac{\beta G \sin(G z)\textbf{z}}{\sqrt{1 + \beta^2 G^2 \sin^2(G z)}}.
\end{eqnarray}
Up to second order, the boundary conditions generate terms of the form
\begin{eqnarray}\nonumber
\lefteqn{
\phi_i = A - \frac{2 \lambda}{\epsilon_i}\log r + \beta B I_0(G r)\cos(G z)} \\ &&
 + \beta^2 \bigg \{ C + D I_0(2 G r) \cos(2Gz) \bigg \},
\end{eqnarray}
and
\begin{eqnarray}\nonumber
\lefteqn{
\phi_o = E K_0(\gamma r) + \beta F K_0((G^2 + \gamma^2)^{1/2}r)\cos(Gz)} \\&& + \nonumber
 \beta^2 \bigg \{ G K_0(\gamma r) + H  K_0((4G^2 + \gamma^2)^{1/2}r)\cos(2Gz)  \bigg \},\\
\end{eqnarray}
where $\mbox{I}_0$ is the modified Bessel function of the first kind.  The boundary condition equations further enable one to obtain a system of linear equations which specify the unique solution for the coefficients $A$ through $H$.  The algebra is most easily carried out in a software package such as \textit{Mathematica}.  The resulting expressions are easy to obtain but are too lengthy to report here.  Once the coefficients for the potential have been determined, one can read off the effective charge of the near-cylinder, defined here to be $(E + \beta^2 G)/2$.  This is proportional to the coefficient of $K_0(\gamma r)$ above, a natural choice given that the potential due a screened line of charge is given by $2 \lambda K_0(\gamma r)$, from Eq.~(\ref{oneds}).

\begin{figure}\scalebox{0.80}
{\includegraphics{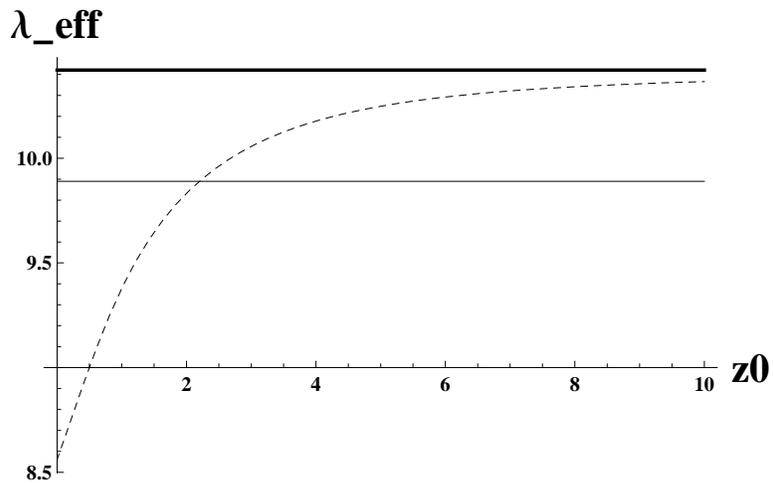}}
\caption{\label{fig:condwave} A plot of the effective charge density versus modulation wavelength for the near-cylinder.  The thin line is the unmodulated charge density, the dashed line is the modulated effective charge valid to second order in $\beta$, and the thick line is the radii-averaged value discussed in the text. The parameter values were set to $\beta = 0.1$, $\epsilon_i = 1$, $\epsilon_0 = 5$, and $\gamma = 5$.}
\end{figure}

We now examine how the resulting effective charge varies with the wavelength of the surface modulation.  At long wavelength the surface curvature is low and the cylinder has a slowly varying local radius.  The effective charge at long wavelength is thus expected to be that which one would obtain through an appropriate average over radii of the effective charge for an unmodulated cylinder.  The appropriate average is
\begin{eqnarray}
\bar{\lambda} = \int_0^1 \frac{du}{\gamma \epsilon_o (1+\beta \cos 2 \pi  u) K_1[\gamma  (1+\beta \cos 2 \pi  u)]}.
\end{eqnarray}
This integral is over a convex function, which, by Jenson's inequality, implies that the modulation will result in an increase rather than a decrease in the effective charge of the cylinder.  A plot of the effective charge, valid to second order in $\beta$, versus wavelength is shown in Fig.~\ref{fig:condwave}.  The effective charge does approach the radii-averaged unmodulated value at long wavelengths, as expected.  Note also that the effective charge decreases as the modulation wavelength decreases, and below wavelength values of $l_0 \approx 2$, the resulting condensation works to reduce the net effective charge of the cylinder.  However, the perturbation expansion converges most quickly when the modulation wavelength is large.  Indeed, the expansion will break down when $\beta G>1$ owing to the fact that the $\beta$ power series of $\hat{\textbf{n}}$ in Eq.~(\ref{nhateqn}) will diverge in this limit.  Thus, convergence concerns indicate that the plot should not be considered quantitatively accurate at small values of $l_0$.  A decrease in the effective charge is also observed at short modulation wavelengths for very small $\beta$, however, where the second order solution is likely to be very accurate.  It is thus reasonable to expect the observed decrease in effective charge at small modulation wavelengths to be qualitatively accurate at larger $\beta$ values as well.  

These two examples demonstrate that the sign of surface modulation-induced condensation can be sensitive to both counterion valence and modulation wavelength.  A modulation wavelength dependence was observed only in the second example but it also holds true for some other systems, linear geometries with constant potential boundary conditions providing one example.  We have focused above on the net condensation for these systems.  It is worth noting, however, that the oscillatory components of these potentials can introduce attractive terms to the interaction between two shifted, identical molecules, just as in the charge modulated example presented earlier.

\section{Discussion}

In this paper, a collection of simple examples have been presented which demonstrate how inherent modulations can alter the potentials within electrolyte solutions.   It was first shown that charge modulation can result in like-charge attraction between identical molecules within the P-B formalism.  The purpose of including this discussion was only to make the technical point that attraction is possible even when the free screening charges are considered at the mean-field level, contrary to some suggestions in the literature.  Next, it was demonstrated that charge modulation results in extra screening charge condensation when the P-B equation is expanded to second order in the potential.  This result, when taken together with the results of Ref.~\cite{Luk:02}, implies that charge modulation-induced condensation universally occurs within the P-B formalism when the potential is expanded to second order in the modulation amplitudes.  In particular, the result applies to systems which have large modulations but little net charge.  In systems such as these, charge modulations can easily result in overcharging.  It was further shown that screening charges of higher valence allow for a reduction in the net cancellation of modulation-induced condensation which occurs in neutral, univalent electrolyte systems.  This is consistent with the fact that overcharging has only been observed in systems which contain multivalent counterions.  Similar results were shown to apply for surface modulated systems and related effects can be expected whenever non-linearities appear in either the differential equation or the boundary conditions which determine the potential.

We conclude by elaborating upon how one can test whether charge modulation significantly contributes to any experimental observation of overcharging.  In the above we noted that mobile charges of equal valence but opposite sign are expected to condense in equal numbers onto a substrate due to inherent charge modulations.  This is not the case for fluctuation-induced condensation, however, where charges are only expected to condense onto substrates of the opposite sign.  This suggests that an overcharging observation can be identified as modulation-induced if the overcharging is negated through the introduction of equal quantities of multivalent counterions of each sign into the system.  We note that this test would not apply to systems where small surface modulations cause condensation in the weak potential limit, where the linearized P-B equation applies.  It would, however, also apply to systems where surface modulations bring about condensation at higher orders in the P-B expansion.  

\subsection*{Note added to manuscript}
Following the publication of this article, I became aware of two references relevant to the included discussion regarding interactions between heterogeneously charged surfaces \cite{Mik:94,Bre:08}.  Both of these predate the present article, and correctly indicate that charge modulation can allow for attractions between like-charged surfaces. 

\begin{acknowledgments}
The author thanks Professors Robijn Bruinsma, Alex Levine, and Giovanni Zocchi for helpful comments and Professor Joseph Rudnick for many helpful discussions.
\end{acknowledgments}

\bibliography{refs}

\end{document}